\documentclass[sigconf, authorversion]{acmart}

\usepackage{makecell}
\usepackage{multirow}
\AtBeginDocument{%
  }

\usepackage[table]{xcolor}

\definecolor{BarGreen}{RGB}{0,179,0}
\definecolor{BarGray}{RGB}{179,179,179}
\definecolor{BarRed}{RGB}{255,51,51}

\newlength{\StackedBarWidth}
\newlength{\StackedBarHeight}
\newcommand{\stackedbar}[3]{%
  \begingroup
  \count0=\numexpr #1+#2+#3\relax
  \ifnum\count0>0
    \dimen0=\dimexpr \StackedBarWidth*#1/\count0\relax
    \dimen2=\dimexpr \StackedBarWidth*#2/\count0\relax
    \dimen4=\dimexpr \StackedBarWidth-\dimen0-\dimen2\relax
    \setlength{\fboxsep}{0pt}%
    \mbox{%
      \raisebox{-0.7ex}{%
        \fbox{%
          {\color{BarGreen}\rule{\dimen0}{\StackedBarHeight}}%
          {\color{BarGray}\rule{\dimen2}{\StackedBarHeight}}%
          {\color{BarRed}\rule{\dimen4}{\StackedBarHeight}}%
        }%
      }%
      \hspace{0.35em}{\tiny #1/#2/#3}%
    }%
  \else
    {\tiny ---}%
  \fi
  \endgroup
}

\copyrightyear{2026}
\acmYear{2026}
\setcopyright{cc}
\setcctype{by}
\acmConference[ITiCSE 2026]{Proceedings of the 31st ACM Conference on Innovation and Technology in Computer Science Education V. 1}{July 10--15, 2026}{Madrid, Spain}
\acmBooktitle{Proceedings of the 31st ACM Conference on Innovation and Technology in Computer Science Education V. 1 (ITiCSE 2026), July 10--15, 2026, Madrid, Spain}
\acmDOI{10.1145/3803400.3809397}
\acmISBN{979-8-4007-2634-7/2026/07}

\begin{document}

\title[Say What? Examining Text and Voice Input Modalities]{Say What? Examining Text and Voice Input Modalities for Prompt-Based Programming in Computing Education}

\author{Kaitlin Riegel}
\orcid{0000-0002-8187-2016}
\affiliation{
  \institution{University of Auckland}
  \city{Auckland}
  \country{New Zealand}}
\email{kaitlin.riegel@auckland.ac.nz}

  \author{Yan Cathy Hua}
\orcid{0000-0001-9155-9667}
\affiliation{
  \institution{University of Auckland}
  \city{Auckland}
  \country{New Zealand}}
\email{yhua219@aucklanduni.ac.nz}

\author{Paul Denny}
\orcid{0000-0002-5150-9806}
\affiliation{
  \institution{University of Auckland}
  \city{Auckland}
  \country{New Zealand}}
\email{paul@cs.auckland.ac.nz}

  \author{Victor-Alexandru Pădurean}
\orcid{0009-0004-2998-096X}
\affiliation{
  \institution{MPI-SWS}
  \city{Saarbrücken}
  \country{Germany}}
  \email{vpadurea@mpi-sws.org}

  \author{Juho Leinonen}
\orcid{0000-0001-6829-9449}
\affiliation{
  \institution{Aalto University}
  \city{Espoo}
  \country{Finland}}
\email{juho.2.leinonen@aalto.fi}

\author{James Prather}
\orcid{0000-0003-2807-6042}
\affiliation{
  \institution{Abilene Christian University}
  \city{Abilene}
  \state{TX}
  \country{USA}
}
\email{james.prather@acu.edu}

\author{Adish Singla}
\orcid{0000-0001-9922-0668}
\affiliation{
  \institution{MPI-SWS}
  \city{Saarbrücken}
  \country{Germany}}
\email{adishs@mpi-sws.org}

\renewcommand{\shortauthors}{Kaitlin Riegel et al.}

\begin{abstract}
Large language models (LLMs) are increasingly integrated into computing education, yet nearly all prior research has focused on text-based interactions. As voice-enabled interfaces become more capable and more common, there is growing interest in understanding how voice input might shape students’ use of LLM-powered tools. In this exploratory study, we investigated how introductory programming students interact with Prompt Problems, which are programming tasks that require crafting natural-language prompts to generate correct code. Students (\textit{N} = 919) solved a series of Prompt Problems with the freedom to select or switch between text and voice input modalities. We collected their prompt submissions as well as post-activity survey responses, then analysed differences in prompt accuracy, persistence, and perspectives by modality. For two of the three problems, we found that students who typed their prompts using text were more likely to have those prompts succeed on the first attempt than students who submitted unedited voice prompts.
There was no difference in success rate if students edited their transcribed voice prompts before submission. Across the problems, we found evidence that students who tried voice prompting varied in their usage of modality – perhaps indicating a complementary, or non-preferential approach. However, most students only tried and reported preferring text. Our qualitative analysis revealed how students’ perceived the roles of voice and text input in shaping their problem-solving process, as well as the reported drawbacks and advantages of each modality. We discuss implications for future multimodal tools and instructional design in computing education.

\end{abstract}



\maketitle

\section{Introduction}

Voice-based assistants have become increasingly capable, evolving from simple command tools to conversational systems that can understand context and support a wide range of tasks \cite{jampala2024evolution}. Modern LLM-powered voice interfaces provide flexible, natural interactions \cite{mahmood2025user, dong2023towards}, and research in education 
shows that voice-based tools can increase motivation and emotional engagement compared to text-based systems \cite{mele2025talking}. There is also emerging evidence that voice input may reduce cognitive load in certain programming contexts \cite{chandu2025voice}. These developments suggest that voice interaction may offer benefits across a range of activities in computing education.

LLM-powered digital teaching assistants have now been widely explored as a way to provide students with timely help and guidance. These tools are used to clarify concepts, debug code, and work through problem-solving steps, and students appreciate their on-demand support and configurable guardrails \cite{denny2024desirable, sheese2024patterns}.  Other LLM-based programming assistants further support comprehension, error diagnosis, and task completion \cite{kazemitabaar2024codeaid, pirzado2024navigating, li2024progmate, lyu2024evaluating, DBLP:conf/aied/PhungCWSB25}. Despite this progress, nearly all existing work in computing education has focused on \emph{text}-based interactions.  Research on programming assistants and question–answering tools has overwhelmingly examined typed prompts \cite{kazemitabaar2024codeaid, sheese2024patterns, pirzado2024navigating, li2024progmate, lyu2024evaluating, liu2024teaching}. Only recently has the field begun to explore voice-based interaction. Jacobs and Kiesler \cite{jacobs2025genai} studied a real-time voice-enabled GenAI tutor and found that, while voice input can offer hands-free, accessible interaction, it also introduces challenges -- in particular, poor verbalisation of code.  These results suggest voice-based interaction may be well suited to conversational or conceptual tasks, but activities requiring precise expression of code elements are more problematic.

A natural next step, then, is to examine voice-based interaction in programming activities that focus on natural language rather than code syntax.  
In a Prompt Problem \cite{denny2024prompt}, students create a natural-language prompt designed to make an LLM generate correct code for a small computational task. Because students express their problem-solving approach directly through language, these tasks offer a clear way to compare how input modalities might influence prompt effectiveness. To date, all existing work on Prompt Problems has used only typed input \cite{padurean2025prompt, padurean2025interleaving,prather2025breaking,denny2024prompt,kerslake2024integrating}

As voice interfaces become more commonly used, it is important to understand how they may shape students' engagement with programming tasks. Voice may change the process of problem solving and students may have different reasons for preferring when and how to use voice or text input.  We therefore take an exploratory approach, examining how students use voice and text, how these choices relate to prompting outcomes, and how students experience each input mode.  
We investigated a series of Prompt Problems, where students were free to choose their preferred input mode.  We collected information on the accuracy and usage of prompting by modality and surveyed the students about their choices and perspectives. Our study is guided by three research questions:

\begin{figure*}
\includegraphics[width=0.98\linewidth]{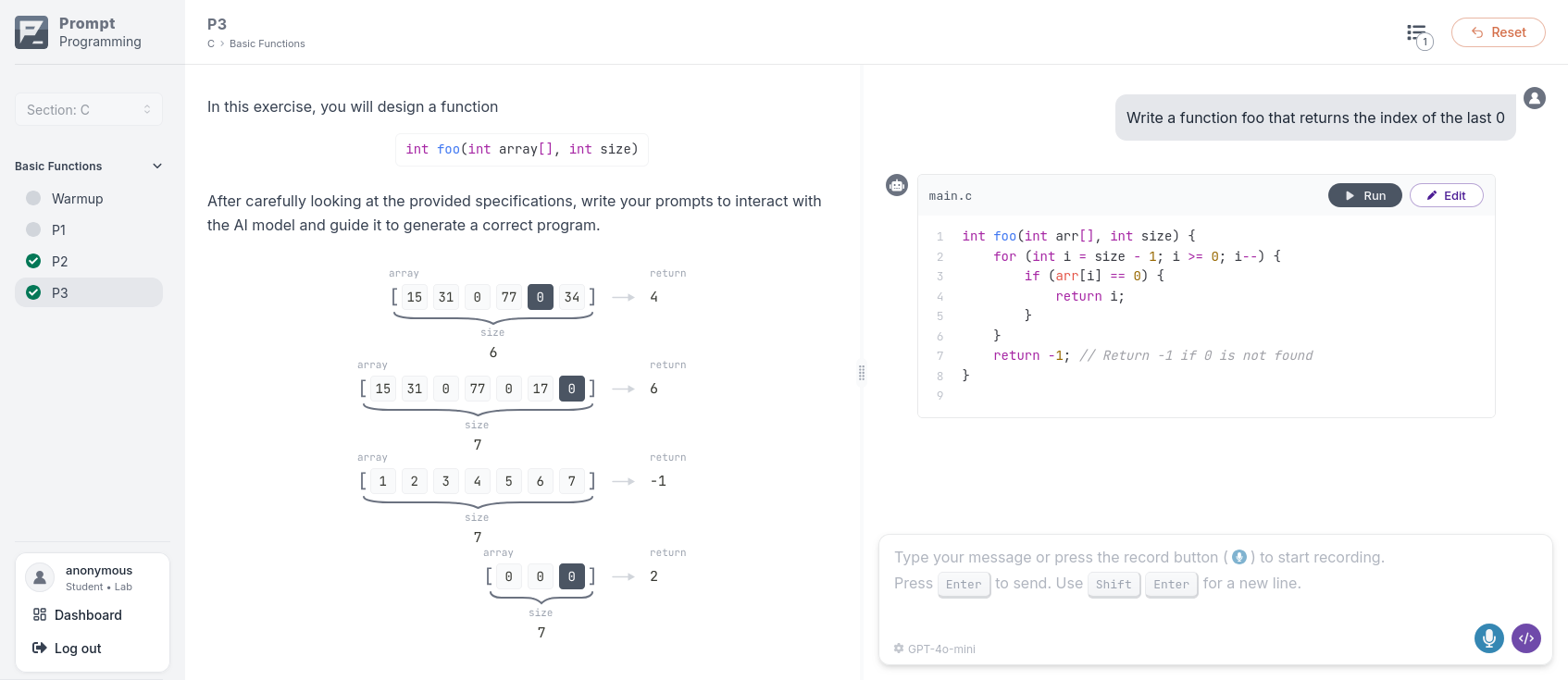}
\caption{\looseness-1The Prompt Programming platform: the Prompt Problem (left), including the required function signature and visual input-output examples, and the chat pane and highlighted code snippets (right), with controls to edit and run code against tests.}
\label{fig:platform}
\end{figure*}

\begin{itemize}
    \item[\textbf{RQ1}:] To what extent does the accuracy of initial prompts submitted using each modality differ? 
    \item[\textbf{RQ2}:] To what extent do students persist in the use of voice prompting, where attempted? 
    \item[\textbf{RQ3}:] How do students' perceptions of each prompting modality differ?
\end{itemize}

\section{Related Work}

Prompt Problems are a natural language programming task where the learner is presented with a visual problem description and their task is to write a prompt for an AI model to generate the code to solve the problem~\cite{denny2024prompt}. The generated code can be run against instructor-defined unit tests to evaluate correctness. Several platforms now support Prompt Problems for classroom use, e.g., \cite{denny2023promptly, padurean2025prompt}. Evaluations of Prompt Problems have found that students enjoy solving them~\cite{denny2024prompt}, that performance demonstrates a weak correlation with code writing (suggesting these possibly target distinct skills)~\cite{kerslake2024integrating}, and that Prompt Problems can support multilingual teaching, as they can be solved in students' native languages~\cite{prather2025breaking}.

All prior work on Prompt Problems has used text as the input modality. However, speech can be up to three times faster than typing as an input modality for text entry on mobile devices~\cite{ruan2016speech}, suggesting improved usability. Rzepka et al.~\cite{rzepka2022voice} argue that speech is more natural and intuitive as an input modality compared to text. They found participants using speech input for an information search task reported higher perceived efficiency, lower cognitive effort, higher enjoyment, and higher service satisfaction, but results were dependent on the task's goal-directness. In the context of educational robots, Mele et al.~\cite{mele2025talking} found that voice modality enhanced emotional and cognitive engagement for students, and improved concentration and emotional connection, while the text modality was preferred for supporting review of content. Thus, they argue that the modalities can be complementary and have different benefits depending on the task. Korkmaz et al.~\cite{korkmaz2024talk} evaluated both voice and text modality as input and output using a 2x2 study design, where each participant experienced all four combinations of input-output modality pairs. They found that participants preferred usability over efficiency. For example, their results suggested that voice input and text output can be very efficient but less preferred, due to lower perceived user experience. Based on their results, the greatest preference was for the text input and output combination, due to giving users a high degree of control and freedom. 

In a programming context, Chandu et al.~\cite{chandu2025voice} found a voice-controlled programming assistant reduced reported typing fatigue and helped debugging. Jacobs and Kiesler~\cite{jacobs2025genai} had ninth grade students use a voice-controlled AI. They similarly found that students mostly used the AI for debugging and perceived it as competent, even though incorrect feedback was given $\sim$30\% of the time. They also found a major problem was poor verbalisation of programming constructs in the voice modality, sometimes leading to incorrect or nonsensical outputs. To the best of our knowledge, Jacobs and Kiesler present the only study in programming education that examines voice as an input modality for generative AI–based assistance. Our work complements this prior study by examining voice as an input modality for Prompt Problems and analysing how students’ modality choices relate to outcomes and perceptions.

\section{Methods}

\paragraph{Platform}
Our study employed the publicly available Prompt Programming web platform \cite{padurean2025prompt} (see Figure~\ref{fig:platform}). 
The problems we used were selected from the library of problems available on the platform, and we collected interaction logs, which were anonymised prior to analysis. When working on the problems, students could submit prompts either by typing or by recording speech via an in-browser microphone control. Voice recordings were transcribed via the OpenAI API using whisper-1, powered by the open-source Whisper V2 model \cite{introducingwhisper,whisperdeveloper}; prior work reports English benchmark word error rates below 5\% \cite{DBLP:conf/icml/RadfordKXBMS23}. The transcription request did not include a domain-specific prompt indicating that computing-related terminology should be expected. The resulting transcript was shown in the message box before submission, and students could either send it as-is (\textit{unedited voice}) or edit it before sending (\textit{edited voice}). The GPT-4o-mini model was used to support the chat assistant, with a system prompt directing it to return only task-relevant code in the requested language and format, without extraneous boilerplate.
Interaction logs contained the sent messages, the transcribed voice recordings, the model responses, and code execution results. 

\paragraph{Course context and tasks}

The study was run in an introductory C programming course at the University of Auckland in Semester Two of 2025 and data analysis was approved by the University's Human Participants Ethics Committee (\#25279). On one of the weekly labs, we configured four problems on the platform: summing two given arguments (\textbf{warm-up}); counting negative values in a given array (\textbf{P1}); summing even values in a given array (\textbf{P2}); and returning the index of the last zero in a given array (\textbf{P3}). Following the tasks, students responded to two reflections: ``What combination of `voice' input and `text' input did you find was most effective when working on the problems?'' (options: \textit{Voice input only}, \textit{Mostly voice input, with a little text input}, \textit{An equal mix of voice and text input}, \textit{Mostly text input, with a little voice input}, and \textit{Text input only}); and ``Please comment on your experience using the different input modes (`voice' input and `text' input) to solve these `prompt programming' tasks''. 1038 students were enrolled in the course. 52 did not engage with the lab or the reflections and 55 engaged after the deadline, so were excluded from the analysis. 13 students did not engage in any prompting and were also excluded. Five students only attempted the warm-up.

\paragraph{Prompt analysis}
Binary logistic regressions were conducted across \textbf{P1-3} to examine the effect of initial prompt method (\textit{unedited voice}, \textit{edited voice}, or \textit{text}) on immediate success (i.e., if the code generated in the first model response succeeded), where 0 = failure and 1 = success. The model was estimated using maximum likelihood, and model significance was assessed with the likelihood-ratio omnibus test. Individual predictor effects were evaluated using Wald $\chi^{2}$ statistics. As this study is exploratory, uncorrected \textit{p} values are reported. 
For students who attempted both prompting approaches, a generalised linear mixed model was employed to examine the influence of students' previous input choice on their next choice (including the warm-up problem). A binomial distribution with a logit link function was specified. The model included fixed effects for previous choice and a random intercept for each participant.

\paragraph{Reflection question analysis}

To analyse responses from students to the open-ended question, we adopted a multi-label classification approach that assigns zero or more predefined category and sentiment labels to each response text entry. We obtained the category labels using a combination of AI and manual processes. We used two pre-trained LLMs (Gemini 3.0 Pro\footnote{\url{https://deepmind.google/models/gemini/pro/}, accessed via an institutional licence} and GPT-5\footnote{\url{https://openai.com/index/introducing-gpt-5/}, accessed via \href{https://copilot.microsoft.com/}{Microsoft Copilot} with institutional licence}) to summarise key topics from all response text entries as the initial category labels. We then iteratively repeated the following steps to refine the labels: 1) manually review and modify the labels to align with our research focus; 2) use the LLM to assign the labels to each entry; 3) review the label assignments and the distribution of entries across labels to evaluate category independence and fit for the data; and 4) manually update the labels to begin the next iteration.

To assign category and sentiment labels to each text entry, we leveraged the Aspect-based Sentiment Analysis (ABSA) method, which extracts all text segments relevant to the provided categories and assigns category and sentiment labels to each extracted segment \cite{liu2012_sentimentanalysisbook, hua2024_absareview}. For example, ``\emph{I like using the text input most, because I can control it very easy}" 
extracted aspect ``\emph{text input}" with opinion ``\emph{can control it very easy}," and was assigned category \textit{Input Accuracy and Control (Text)} with a \textit{positive} sentiment. To speed up the process, we performed the initial segment extraction and label assignment using a custom prompt with an ABSA-fine-tuned small LLM from \citet{hua2025_edurabsaslm}\footnote{We used the Phi4-mini version from \url{https://huggingface.co/yhua219/EduRABSA_SLM_v1_SLERP_phi4mini}}, which showed better opinion extraction spans on a small pilot dataset compared to pre-trained LLMs. Two human annotators then independently re-annotated 250 randomly selected review entries from the LLM output file. Within the 250 human-annotated entries, the first 94 were used as a pilot to develop the annotation rules through the difference-resolution process, and the independent annotations of the remaining 156 entries were used to calculate the inter-rater reliability. Inter-rater agreement was moderate (micro-averaged $\textit{F}_1$ = 0.65), reflecting the interpretive and multi-label nature of the task. Disagreements were subsequently resolved through discussion to produce a consensus-coded dataset for all 250 entries used for all further analyses. Together, the two researchers spent approximately 45 combined hours on the analysis. For the purposes of this paper, we focus on the categories specifically reflecting on the input modalities.

\section{Results}

\begin{table*}[]
\small
\caption{Descriptive statistics by problem and modality.}
\label{tab_descriptivestats}
\begin{tabular}{llccccc}
\hline
   & \textbf{}      & \multicolumn{1}{l}{\textbf{}} & \multicolumn{1}{l}{} & \multicolumn{3}{c}{\textbf{\textit{M} (SD)}}                                                       \\ \cline{5-7} 
   &                \textbf{Input modality}& \textbf{\textit{N} (attempt)}& \textbf{\textit{N} (success)}& \textbf{Proportion Immediate Success}& \textbf{Messages until Success}& \textbf{First Message Characters}\\ \hline
   & Unedited Voice & 51                            & 48                   & 0.35 (0.48)                  & 1.69 (1.60)            & 221.41 (107.07)          \\
\textbf{P1}& Edited Voice   & 26                            & 25                   & 0.38 (0.50)                  & 1.56 (0.96)            & 205.00 (113.90)          \\
   & Text           & 830                           & 824                  & 0.52 (0.50)                   & 1.71 (1.89)            & 183.17 (96.47)           \\ \cline{3-3}
   &                & Total \textit{N} = 907                 &                      &                              &                        &                          \\ \hline
   & Unedited Voice & 40                            & 39                   & 0.38 (0.49)                  & 1.67 (1.16)            & 206.15 (100.23)          \\
\textbf{P2}& Edited Voice   & 30                            & 30                   & 0.50 (0.51)                  & 1.23 (0.43)            & 197.07 (76.33)           \\
   & Text           & 831                           & 829                  & 0.58 (0.49)                  & 1.53 (1.35)            & 193.41 (100.28)          \\ \cline{3-3}
   &                & Total \textit{N} = 901                 &                      &                              &                        &                          \\ \hline
   & Unedited Voice & 35                            & 34                   & 0.31 (0.47)                  & 2.15 (2.16)            & 272.80 (107.31)          \\
\textbf{P3}& Edited Voice   & 35                            & 33                   & 0.26 (0.44)                  & 2.76 (2.96)            & 271.52 (186.95)          \\
   & Text           & 830                           & 816                  & 0.36 (0.48)                  & 2.47 (2.78)            & 226.62 (118.82)          \\ \cline{3-3}
   &                & Total \textit{N} = 900                 &                      &                              &                        &                          \\ \hline
\end{tabular}
\end{table*}

\subsection{Initial Prompt Success by Initial Modality (\textbf{RQ1})}
Table \ref{tab_descriptivestats} presents descriptive statistics by problem and input modality (i.e., \textit{unedited voice}, \textit{edited voice}, and \textit{text}). We employed binary logistic regression models across each of the three problems to examine how the different approaches to constructing initial prompts influenced students' immediate success. For both \textbf{P1} and \textbf{P2}, the overall models were significant ($\chi_{P1}^2$ = 7.405, \textit{p} = .025; $\chi_{P2}^2$ = 7.158, \textit{p} = .028). Compared to \textit{text} prompting, students who initially used \textit{unedited} voice prompting had lower odds of immediate success (OR$_{P1}$ = 0.50, 95\% CI [0.27, 0.89], \textit{p} = 0.02; OR$_P2$ = 0.43, 95\% CI [0.23, 0.83], \textit{p} = 0.01), while students who \textit{edited} their voice prompts did not significantly differ (OR$_P1$ = 0.57, 95\% CI [0.26, 1.27], \textit{p} = 0.17; OR$_P2$ = 0.72, 95\% CI [0.35, 1.49], \textit{p} = 0.38). In \textbf{P3}, the model was not significant, $\chi^2$ = 1.948, \textit{p} = .378, indicating initial prompting method did not explain immediate success. Together, the results suggest the possibility that students who edit their voice prompts are as likely to succeed as students using text. However, unedited voice prompting may be unreliable.  

\subsection{Persistence of Voice Prompting ({\textbf{RQ2})}}
Including the warm-up problem, there were 813 students who only used text-based prompting (88.5\%), 44 who only used voice (4.8\%), and 62 who attempted both input methods (6.7\%), demonstrating an overwhelming bias towards text prompting. However, we were interested in the influence of the novel voice input on students' behaviours. Consequently, for the 62 students who attempted both prompt approaches, a binary, mixed-effects logistic regression was conducted to examine whether previous prompt modality influenced subsequent modality selection. The effect was not statistically significant (OR = 1.28, 95\% CI [0.71, 2.31], \textit{p} = 0.41), indicating that, for this subset, students' choices were independent of their previous choices. In addition to the 44 students who used voice and continued to only use voice for their initial prompts, the absence of a significant effect of voice prompting on subsequent modality use suggests that students are not deterred by voice-based interactions. A possible interpretation is that students flexibly combine voice and text prompting according to their needs.

We sought to examine the persistence of using voice prompts within problems, however, as many students immediately succeeded on the problems, we did not attain sufficient sample sizes. For the small subsets who initially engaged with voice prompting and did not experience immediate success, we found there were some who continued using voice prompting within the problem. No strong conclusions can be drawn, but the results could hint that some students engage with voice prompting in a dialogue-based capacity. 

\begin{figure}
\includegraphics[width=\linewidth]{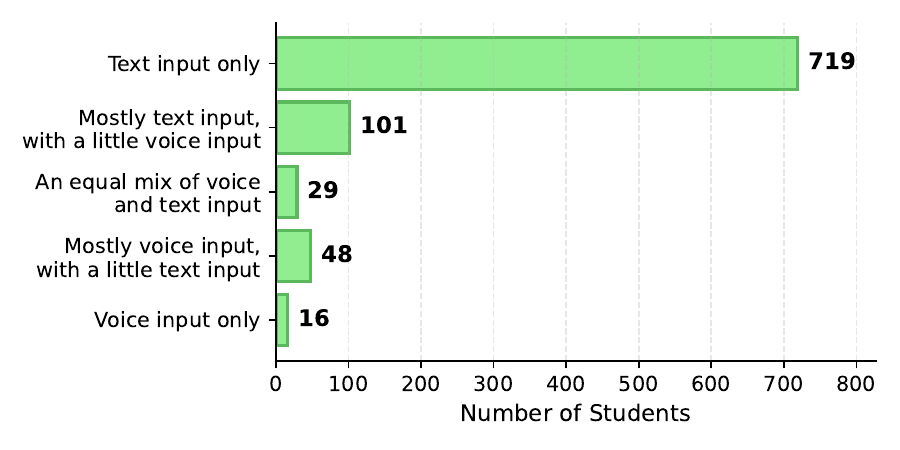}
\caption{Student input modality preferences.}
\label{fig_preferences}
\end{figure}

\subsection{Students' Perceptions of Prompting Modalities (\textbf{RQ3})}
Figure \ref{fig_preferences} summarises responses to the reflection question on mode preference and reveals students overwhelmingly preferred only text. However, the log data showed few opted to attempt using voice input. 
For this reason, we found it appropriate 
to examine the relationship between the proportional usage of the other modalities (\textit{edited voice} and \textit{unedited voice}) with reported preference (measured on a five-point scale). A Spearman rank-order correlation was conducted between preference (\textit{``Text input only''} = 1 and \textit{``Voice input only''} = 5) and proportion of both \textit{edited} and \textit{unedited} voice prompts. In both cases, there was a moderate, positive correlation between a greater proportion of voice prompts and a greater preference for using voice ($\rho_{edited}$ = .509, \textit{p} < .001 and $\rho_{unedited}$ = .597, \textit{p} <.001).  Consequently, we received insight that students who \textit{do} engage in voice prompting are more likely to enjoy it -- suggesting, again, they are not being deterred by the modality.

\begin{table*}[] 
\footnotesize 
\caption{Category labels and definitions with respect to prompt input modality. Counts for \textit{Positive}, \textit{Neutral}, and \textit{Negative} sentiments within each category represented in the bars as green, grey, and red, respectively.} 
\label{table_category_definition} 
\begin{tabular}{p{3cm}|p{6.5cm}|p{3.5cm}|p{3.5cm}} 
\hline 
\textbf{Category Label} & \textbf{Definition} & \textbf{Text Input}& \textbf{Voice Input}\\ 
\hline 
Affective Response & Sentiment of the input mode with no further details & 
\stackedbar{33}{3}{0} & 
\stackedbar{11}{1}{12} \\ 

Ease of Use & How easy or hard it is to use the input mode & 
\stackedbar{26}{0}{1} & 
\stackedbar{7}{0}{2} \\ 

Efficiency \& Speed & Speed of input mode execution or the overall process involving the input mode & 
\stackedbar{13}{0}{5} & 
\stackedbar{16}{2}{6} \\ 

Planning \& Thinking Process & How the input mode facilitates or hinders the user's ability to think, plan, structure thoughts, or manage cognitive load & 
\stackedbar{44}{0}{1} & 
\stackedbar{6}{3}{17} \\ 

Editability \& Refinement & The ability to review or edit the prompt after drafting but before execution & 
\stackedbar{47}{0}{2} & 
\stackedbar{0}{0}{12} \\ 

Technical Expressions \& Syntax & Relating to using the input mode for programming / technical tasks & 
\stackedbar{5}{0}{1} & 
\stackedbar{0}{0}{3} \\ 

Input Accuracy \& Control & Relating to the accuracy of transcription errors, input precision, or the user's ability to input prompts precisely and with control & 
\stackedbar{53}{0}{2} & 
\stackedbar{6}{3}{51} \\ 

Familiarity \& User Confidence & User familiarity and/or confidence in using a particular input mode & 
\stackedbar{6}{1}{0} & 
\stackedbar{2}{1}{0} \\ 

Environment & External factors affecting the choice or experience, (e.g., noise levels or social etiquette) & 
\stackedbar{2}{0}{0} & 
\stackedbar{0}{0}{20} \\ 

Hardware Requirements & Hardware or software limitations (e.g., broken microphone) & 
\stackedbar{2}{0}{0} & 
\stackedbar{0}{0}{18} \\ 

Personal Factor & Individual constraints or preferences (e.g., language barriers, accessibility issues, or specific personal situations) & 
\stackedbar{10}{2}{0} & 
\stackedbar{1}{0}{7} \\ 

Voice - Superfluous& General opinion that the voice input mode is unnecessary, since the tasks can be solved with text.& 
\multicolumn{1}{c}{\small n/a} & 
\multicolumn{1}{c}{\small \textit{n} = 17} \\

Combining Input Modes& Opinions about using both input modes for different aspects of the task & 
\multicolumn{2}{c}{\stackedbar{11}{6}{1}} \\

\noalign{\vskip 2.5pt}

\hline 
\end{tabular} 
\end{table*}

Table \ref{table_category_definition} presents the results of the qualitative analysis, including the category-sentiment labels, definitions, and counts. The initial LLM-proposed categories largely mapped to the final Text/Voice-specific categories with rewording (e.g., ``Code-Specific Suitability'' to ``Technical Expressions \& Syntax''). From examining the pilot comments, we further added two higher-level categories, ``AI/Activity'' and ``Problem Solving,'' each with subcategories, and two input-modality categories, ``Familiarity \& User Confidence'' and ``Personal Factor'', to capture the themes. We discuss and provide student excerpts representing the dominant categories. 

\subsubsection{Input Accuracy and Control}
The dominant category that emerged was the ability to input prompts with control and accuracy, heavily in favour of text-based prompting. One student explained, ``\emph{I find using text input to be my preferred method as I know that whatever I write is exactly what will be passed onto the computer, as with voice it is possible for it to misunderstand what I am saying.}'' Many students cited transcription errors as an issue, including non-native English speakers (NNES). We conducted a post-hoc analysis on all 890 comment pieces to identify those with explicit mention of NNES status. Using a combination of keyword search, LLM categorisation (pre-trained Qwen3.5-4B in non-thinking mode \cite{qwen35blog}), and human consolidation, we identified 17 entries that explicitly mentioned NNES status as a reason for preferring text input. 15 of these comments expressed concerns about their accent being accurately recognised. The other two mentioned uncertainty about voice recognition picking up their native language and time pressure in constructing voice commands in English.

\subsubsection{Editability and Refinement}
The ability to edit and review text prompts before submitting was a major reported advantage. Specifically, ``\emph{When you talk you can make mistakes and you'd have to rerecord the entire audio message if you want to redo it. With typing you can simply backspace.}'' This was an important category because voice-based revision was perceived strictly negatively. Further, the students felt if text-based editing was necessary, they should simply use text for the whole prompting process. 

\subsubsection{Planning and Thinking Process}
A noteworthy finding is the ways the different input methods shape the reported problem-solving process, affecting students' usage decisions. Students frequently referenced the opportunity to structure cohesive thoughts or the ability to engage in non-linear problem-solving using text inputs, exemplified by comments such as, \emph{``Text input allowed me to better collect my thoughts''} and \emph{``I prefer using the text input because I can fully formulate and think through my prompt before actually submitting it''}. One student summarises the conflict between the input methods' influence on their ability to think and plan as follows: 
\begin{quote}
\small
\textit{I am not bottlenecked by my typing speed, but rather, my thought speed...when using voice, I have to think about what to say beforehand, otherwise I will give the AI a confusing and poorly structured prompt...typing things out, I can think on the fly, and edit previous parts of my sentence to refine the prompt. If I am using voice input, then I either have to spend time thinking about what to say, or spend time editing the prompt I gave the AI -- time better spent typing out the prompt.} 
\end{quote} 

In contrast, some students thought voice input served a valuable role in their prompt construction -- often to get initial ideas out: \emph{``Voice input was a lot easier to get down what I was thinking into words as when I was speaking to the program I tended to think more about what the code needed to do than when I wrote it myself.''} 

\subsubsection{Other Categories}
\textit{General Sentiment} and \textit{Ease of Use} were much more positive for text input. Further, the \textit{Voice - Superfluous} category indicated many students default to text and may avoid novelty if traditional methods meet their needs. These results align with students' usage and preferences. However, the analysis highlighted many avoid voice input for practical reasons, specifically, being in an inappropriate environment (\emph{``Speaking to the AI through audio can be especially bad if there is background noise''}, or simply lacking the hardware (\emph{``My laptop doesn't have a microphone.''}).

Interestingly, there was a fairly even distribution for students positively perceiving the efficiency and speed of each input method. One wrote ``\emph{I only used text input, because realistically, it is easier and faster to type out whatever I want to say than dictate, then wait for the programme to transcribe it.''} 
In contrast, another commented, ``\emph{The voice input tended to be much faster than writing out all the text...I find that voice is much better than writing it, especially if you have a lot to write.}'' Finally, some students discussed the interwoven roles of each input modality and how a combination may be the best approach. For example, one stated, ``\emph{I liked starting with voice for the rough idea and then switching to text to clean it up.''} Another commented on the differing contexts in which the modalities are appropriate: ``\emph{I used only text input which worked well for these more simple and short codes. I would be more inclined to use voice prompting for more complicated coding tasks.}''

\section{Discussion}
This study examined students' success, persistence, and perceptions when solving Prompt Problems using voice versus text inputs. Critically, voice engagement did not reduce student preference or use. For those who initially used voice prompts, input modality failing to predict subsequent choices may be due to strategically combining use of text and voice, consistent with prior work \cite{mele2025talking}. 

There was a heavy skew toward students' engagement with and preference for text input, mirroring 
Zavaleta Bernuy
et al. \cite{zavaleta2024does}, who found students preferred text over voice as a medium for self-explanations. 
Our findings appear driven not only by practical barriers (e.g., hardware, environment), but also by low perceived control \cite{pekrun2006control} (e.g., \textit{Voice - Superfluous}, \textit{Text - Familiarity \& User Confidence}) and concerns the negative perception of \textit{Input Control and Accuracy}, aligning with Korkmaz et al. \cite{korkmaz2024talk}. Future work should improve voice usability, to support student control over success, through better hardware, appropriate environments, native language options  \cite{kerslake2024integrating, prather2025breaking}, explicit instruction, and adaptive transcription.

Although students were more likely to experience immediate success using text prompts than \textit{unedited} voice prompts, this may not reflect better learning. Over-reliance on AI tools may cause cognitive functioning to deteriorate \cite{kosmyna2025your}, as well as weaker articulation of solutions \cite{prather2024widening,kazemitabaar2023studying}, poorer outcomes \cite{margulieux2024self}, and diminished metacognition and self-regulation \cite{prather2024widening} in programming education. Thus, faster success may come at the cost of deeper understanding and this could potentially be counteracted using a voice prompting technique.

Moreover, it is critical to evaluate how modality shapes engagement. Students preferred text for its non-linear, iterative planning (\textit{Planning \& Thinking Process}), but this may reduce retention. Information is moved out of the working memory once it is written down \cite{Sweller2011CLT}. Consequently, students can succeed on a task, while focusing only on a single aspect of the problem for a fleeting moment (or with heuristic iterative adjustments) without it sinking into their long-term memory, failing to cause learning. In contrast, voice-based prompting represents all the benefits of speak-aloud self-explanation \cite{bisra2018inducing}. Notably, \textit{edited} voice prompting achieved a similar success rate to text, suggesting voice and text may combine effectiveness with deeper cognitive processing and warrant further investigation.

\section {Limitations and Future Work}

A key limitation is students' self-selection into modality, with few students choosing voice, limiting generalisability of the results and motivating a controlled study. Additionally, causality cannot be established (i.e., whether modality influences outcomes, or student characteristics drive modality choice). Some students, including non-native English speakers, reported transcription issues, suggesting evaluations may reflect technology performance rather than modality. Future studies should examine outcomes when students can voice prompt in their native language. Transcription delay was not measured and may have influenced student behaviour and perceptions. Possible future avenues for this research could include the use of validated surveys measuring technology acceptance \cite{davis1989} and monitoring how students interact with LLMs in non-programming tasks. The tasks in this study were relatively simple and short, which may have not been suitable for comparing modalities. Longitudinal evaluation of retention or potential disadvantages should be explored in future work with more complex tasks and correspondingly longer voice and text prompts.

\section{Conclusions}
This exploratory study illuminated the complex trade-offs between using text and voice input modalities for prompt-based programming. While students showed a clear preference for text, voice input holds promise as a tool for deeper cognitive engagement and self-explanation, with the potential to support long-term learning outcomes. Moreover, immediate success may not translate to meaningful understanding, as over-reliance on AI and superficial strategies can undermine metacognition and retention. Future research into this area should focus on better faciliting the use of voice input and examining its impact in controlled settings. 

\begin{acks}
This work was supported by Research Council of Finland grant \#356114.
\end{acks}

\balance

\bibliographystyle{ACM-Reference-Format}
\bibliography{main}

\end{document}